\documentclass[12pt,manuscript]{aastex}

\title{Lensing Systematics from Space: Modeling PSF effects in the SNAP survey}

\author{H. F. Stabenau$^\dag$, B. Jain$^\dag$, G. Bernstein$^\dag$,
  M. Lampton$^\ddag$}
\affil{$^\dag$ Dept. of Physics and Astronomy, University of Pennsylvania,
Philadelphia, PA\\
$^\ddag$ Lawrence Berkeley National Laboratory, Berkeley, CA}
\email{hstabena, bjain, garyb@physics.upenn.edu, mlampton@ssl.berkeley.edu }
\begin{document}

\begin{abstract}
  Anisotropy in the point spread function (PSF) contributes a
  systematic error to weak lensing measurements. In this study we use
  a ray tracer that incorporates all the optical elements
  of the SNAP telescope to estimate this effect. Misalignments in the
  optics generates PSF anisotropy, which we characterize by its
  ellipticity. The effect of three time varying effects: thermal
  drift, guider jitter, and structural vibration on the PSF are
  estimated for expected parameters of the SNAP telescope.  Multiple
  realizations of a thousand square degree mock survey are then
  generated to include the systematic error pattern induced by these
  effects. We quantify their contribution to the power spectrum of the
  lensing shear. We find that the dominant effect comes from the
  thermal drift, which peaks at angular wavenumbers $l\sim 10^3$, but
  its amplitude is over one order of magnitude smaller than the
  size of the expected statistical error. While there are significant 
  uncertainties in our modeling, our study indicates that time-varying PSFs
  will contribute at a smaller level than statistical
  errors in SNAP's weak lensing measurements. 
\end{abstract}

\section{Introduction}

Current measurements of cosmic shear in the literature have systematic
errors at or below the $\lesssim10\%$ level, most likely generated by
insufficiently corrected optical and atmospheric distortions.  Given
the accuracy requirements for the shear (between 0.1-1\%) for the SNAP
survey \citep{big_SNAP_paper, Rhodes_et_al}, it is clear that progress
on the optical distortions and shape measurement techniques is
essential.  The PSF may have spatial and temporal variations and is
sampled at only the locations of stars, from where it must be
interpolated onto the galaxies.  If each exposure had many stars
distributed around the image, then we wouldn't have to worry about
time variations in the PSF since there would be enough information in
each exposure to subtract the PSF.  If the usable star density is too
low however, subtracting a single static PSF from every exposure will
be inadequate.  Recent work shows that the many exposures used in the
planned SNAP survey could be used to make this correction very
precisely \citep{Ja04}. In this paper, we estimate the shear power
spectra for a fully dynamic treatment of the observatory with thermal
and mechanical PSF disturbances that continue throughout the one year
survey.  To model a first-order PSF correction, we use an imperfect
static PSF that is held fixed throughout the same one year survey.
A further improvement not pursued here would be to update the PSF
model throughout the survey using daily field star images.  Our shear
power spectra should therefore be regarded as upper limits to the
corrected shear spectra, given the model parameters we have assumed.

An estimate of the limiting systematics for ground- and space-based WL
data requires some knowledge of the amplitude and variability of the
PSF.  We are working with other members of the SNAP Weak Lensing
Working Group (WLWG), and the SNAP optical-mechanical engineers, to
model the temporal behavior of its PSF.  In this section we describe
our work on systematic errors: calculation of the expected
instrumental distortion for SNAP by modeling the dominant sources of
PSF anisotropy.

\begin{figure}[t]
\epsscale{1.0}
\plotone{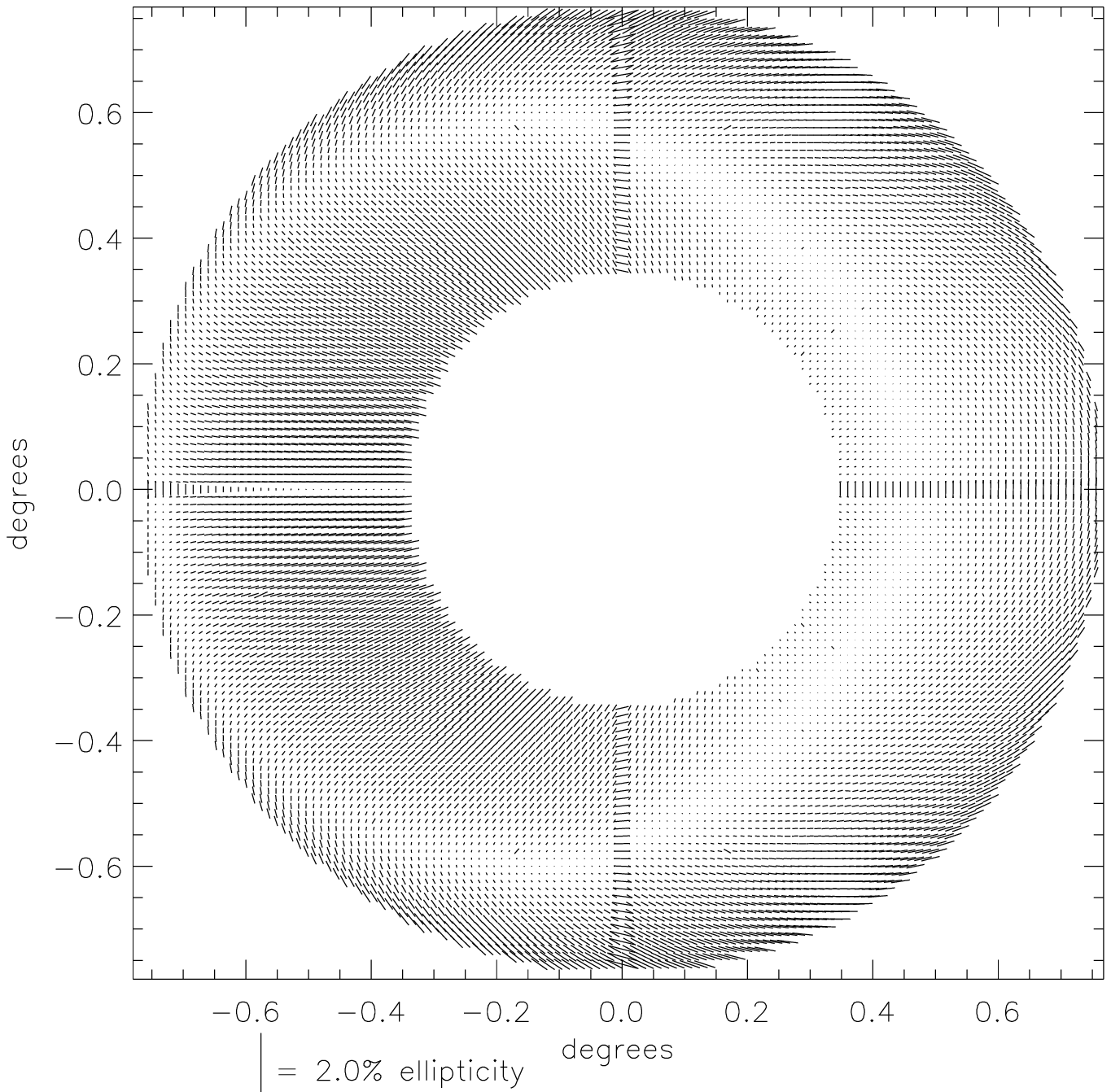}
\caption[]{\small Anisotropy in the point spread function
  (PSF) in the SNAP focal plane. Possible misalignments in the
  telescope optics have been modeled by ray tracing.  In this figure
  we have modeled a large misalignment: the secondary mirror is
  off-center by $10\mu$m.  The PSF at each position is obtained by
  computing the second moment of the intensity. The RMS size of the whiskers
  shown represents a $0.2\%$ ellipticity.  }
\label{whisker_snap}
\end{figure}

PSF anisotropy and time variation have been the dominant systematic
errors in most lensing measurements \citep{Refregier_PSFreview}. This
is one reason why a space instrument is expected to be superior for
lensing.  The expected variations, {\it e.g.}  thermal effects, should
be very slow, given the absence of gravity and the nearly constant
thermal environment.
It is therefore important to quantify the time rate of change of the
PSF for a nominal SNAP mission to determine the degree to which the
recovered shear maps can be corrected for instrumental effects.
An initial assessment of the SNAP PSF time
variations was presented by \citet{Sholl_et_al}.

\section{PSF of the SNAP telescope}

We have calculated the PSF ellipticity on a grid over the field of
view of the instrument, subject to specified optical misalignment
parameters. The first step is to use a 
ray-tracing telescope simulator for SNAP 
to generate monochromatic PSF images at one micron wavelength as a
function of position of the field of view and telescope misalignment.
The ray tracer filled the aperture with a grid
of rays from a distant point source, propagating them through the
optical system to evaluate the pupil's complex transmission, and
then Fourier transforming and squaring to get the image irradiance.
Finally, we determined the second moments of the
image and from them its ellipticity components.

The resulting pattern of PSF anisotropy on the focal plane for a
sample misalignment is shown in Figure~\ref{whisker_snap}. The length
and orientation of each whisker represents the magnitude and
orientation of the ellipticity at that position. The typical
ellipticity shown is $\sim 0.2\%$.  We define the ellipticity in terms
of the second moments of the diffracted light,
\begin{eqnarray*}
  e_1 & = & \frac{{\rm I_{xx}} - {\rm I_{yy}}}{{\rm I_{xx}} + {\rm I_{yy}}} \\
  e_2 & = & \frac{2 {\rm I_{xy}}}{{\rm I_{xx}} + {\rm I_{yy}}}.
\end{eqnarray*}
The second moments ${\rm I_{ab}}$ for $a,b \in \{{\rm x,y}\}$
are defined as
\begin{displaymath}
  {\rm I_{ab}} = \frac{\sum_{j} I_j a_j b_j}{\sum_j I_j},
\end{displaymath}
where $I_j$ are the intensity values of the image pixels, and x$_j$,
y$_j$ are the coordinates of a pixel on the focal plane.  The second
moments ${\rm I_{xx}}$ etc. were computed on an unweighted basis for
this study.  Applying a weighting kernel appropriate for individual
galaxy sizes would slightly reduce our moments but not significantly
influence our conclusions. Whiskers are plotted at an angle of
\[\theta=\frac{1}{2}\arctan{\frac{e_2}{e_1}}\] with respect to the
x-axis, so that
\begin{itemize}
\item pure positive $e_1$ corresponds to a horizontal whisker (--)
\item pure negative $e_1$ corresponds to a vertical whisker (\verb=|=)
\item pure positive $e_2$ corresponds to a 45 deg whisker (/)
\item pure negative $e_2$ corresponds to a 45 deg whisker ($\backslash$)
\end{itemize}

We next consider the process of
generating a survey with SNAP, including effects that cause the PSF
ellipticity pattern to change over the length of the survey.  This
could happen if there were some time-varying instrumental effects,
which would translate into a systematically changing PSF across the
field of the survey.  We modeled a step and repoint survey by taking
an exposure using the chips for one filter band, regenerating the PSF,
moving the telescope by one CCD chip width, and repeating the process,
resulting in a series of strips that we then put together to form a
$32\times 32$ square degree mock survey. Multiple realizations of this
survey were used to compute the power spectrum.

We considered three time-varying misalignment effects when generating
the survey, categorized by the timescale of their variation:
\begin{itemize}
\item Thermal drift, which has a timescale of days to weeks;
\item Telescope structural vibration, which has a timescale of an
 hour; and
\item Telescope guide jitter, which is different for every
 exposure.
\end{itemize}
Of these three effects, the thermal drift effect dominates the
time-varying component of the PSF, while the effect of telescope guide
jitter turned out to have almost no measurable contribution.  
The thermal contributions to the PSF include a daily heat pulse from
daily attitude maneuvers for telemetry, plus long term drifts from
slow changes in the solar longitude and from control system drift.  We
ignore the daily environmental
component because the thermal response is dominated by the thermal control
loop on this time scale.  We attempted to measure the
slower thermal drift contribution to the PSF power spectrum model by
modeling a drift in the control system of the sensors that measure the
temperature of the struts that hold up the secondary mirror, which
undergo thermal expansion and contraction.  If the struts have varying
lengths, the secondary mirror will be defocussed, decentered, and
tilted.

The SNAP focal plane includes dedicated star tracker image sensors to
determine the instantaneous attitude \citep{Lampton_star_camera}.  
By analyzing the positions of
stars within an image and comparing to catalogs, the pointing of the
telescope is established and corrected via a feedback mechanism with
telescope orientation controls.  However Poisson noise makes the
centroids of stars in the image appear to shift around.  
The attitude
control system responds to this centroid jitter according to its
control bandwidth, which cannot be zero.
We modelled the jitter by assuming a continuing stream of star
centroid information from the star guiders, furnished at a rate of 10
frames/second, with an individual centroid RMS noise level of 1
milliarcsecond per axis, as described by \citet{Lampton_star_camera}.
Because these successive centroid determinations are based on
independent photon arrivals, the deviations will be statistically
independent and circularly symmetric around the true long term mean.
We assume that the spacecraft attitude control bandwidth is of the
order of 0.1 Hz, so that $\sim 100$~frames are averaged in establishing each
attitude, and the RMS jitter in each axis will be reduced tenfold from
the individual frame centroids, to $\sim 0.1$ milliarcsecond. The guider's
effect on the PSF turned out to be much smaller than both the
vibrational and thermal effects.

Vibrational modes are excited in the structure of the telescope by the
vibration of four flywheels that store angular momentum used to rotate
the satellite in space.  This causes the secondary mirror to vibrate
with an amplitude which changes as the flywheel goes in and out of
resonance with the telescope structure.  We model the vibration effect
as the sum of two tilted orientations of the secondary mirror, since
the structural vibration modes of the SNAP observatory have
frequencies above 15 Hz, and WL exposures have durations of greater
than 300 seconds, the number of vibration cycles in each exposure is
large,  $> 4500$, allowing us to average the PSF over many cycles.
First, we choose an amplitude for the vibration from a Gaussian
distribution.  Next, we generate two exposures each with the secondary
mirror tilted by plus and minus this amplitude.  Finally, we average
the images from the two orientations.  The result is an exposure
affected by our model vibrational PSF.

Spacecraft systems engineers provided projections of the magnitude of
the thermal and vibrational effects \citep{NASA_report_for_SNAP}.  In
order to simulate surveys affected by time-varying misalignments,
during each model survey new misalignment parameters for the secondary
were chosen from Gaussian distributions every two weeks during the
model surveys.  The widths of the Gaussian distributions used for the
misalignments of the secondary are shown in
Table~\ref{tab:misalignments}.

\begin{table}[t]
  \centering
\begin{tabular}{lcc}
Source & Amplitude & Angular scale \\
thermal decentering along x, y directions & $0.8\mu$m & 7 milliarcsec \\
thermal defocus along z direction & $0.5\mu$m & 4.6 milliarcsec \\
vibrational tilt & $1.6\mu$m & 16 milliarcsec \\
guide jitter & $0.01\mu$m & 0.1 milliarcsec.
\end{tabular}
  \caption{\small Comparison of the RMS widths of the Gaussian
    distributions from which the misalignment parameters are chosen
    during the simulated survey.}
  \label{tab:misalignments}
\end{table}

\begin{figure}[t]
  \epsscale{1.0}
  \plotone{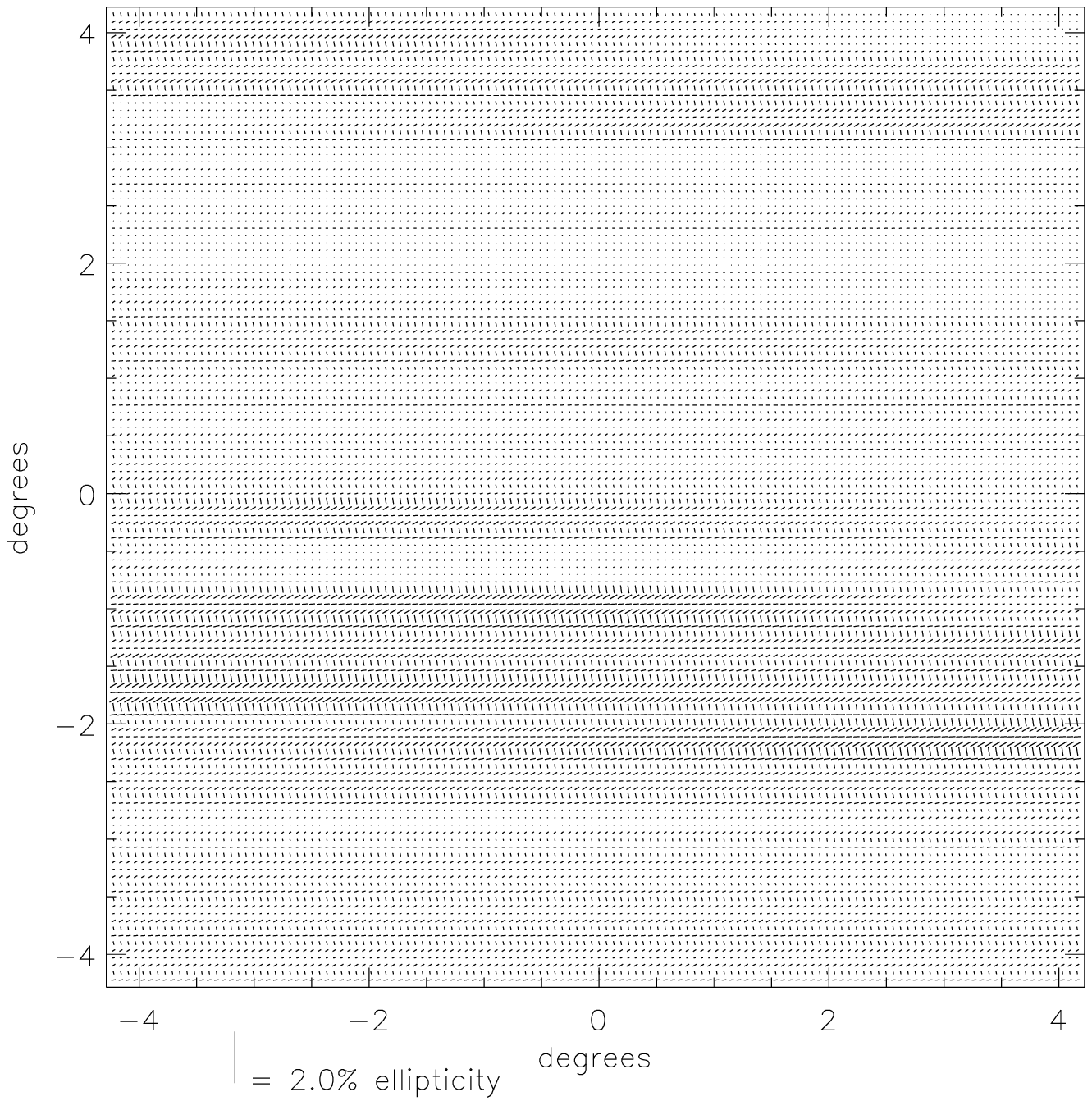}
  \caption{\small PSF anisotropy generated by thermal drift in the
    SNAP telescope. By choosing misalignment parameters from a Gaussian
    distribution every two weeks, we obtain a spatially varying PSF
    pattern with a characteristic length scale.  The whiskers in this
    plot have an RMS ellipticity at the $\sim 0.1\%$ level.  Compare
    with Figure~\ref{rayt_whiskers}, which contains simulated weak
    lensing whiskers, and
    Figure~\ref{compare_resid_rayt}, which plots the power spectra.
    Subtracting off the static pattern produces residuals that are
    smaller by about an order of magnitude, as shown in 
    Figure~\ref{compare_resid_rayt} for the power spectrum.  }
  \label{thermal_whiskers}
\end{figure}

At intermediate times the misalignments
are linearly interpolated between the old and new values.  One such
realization of the set of model surveys can be seen in
Figure~\ref{thermal_whiskers}.  The RMS shear in that figure is about
$0.1\%$.

\section{Shear Power Spectrum}

\begin{figure}[t]
  \epsscale{0.8}
  \plotone{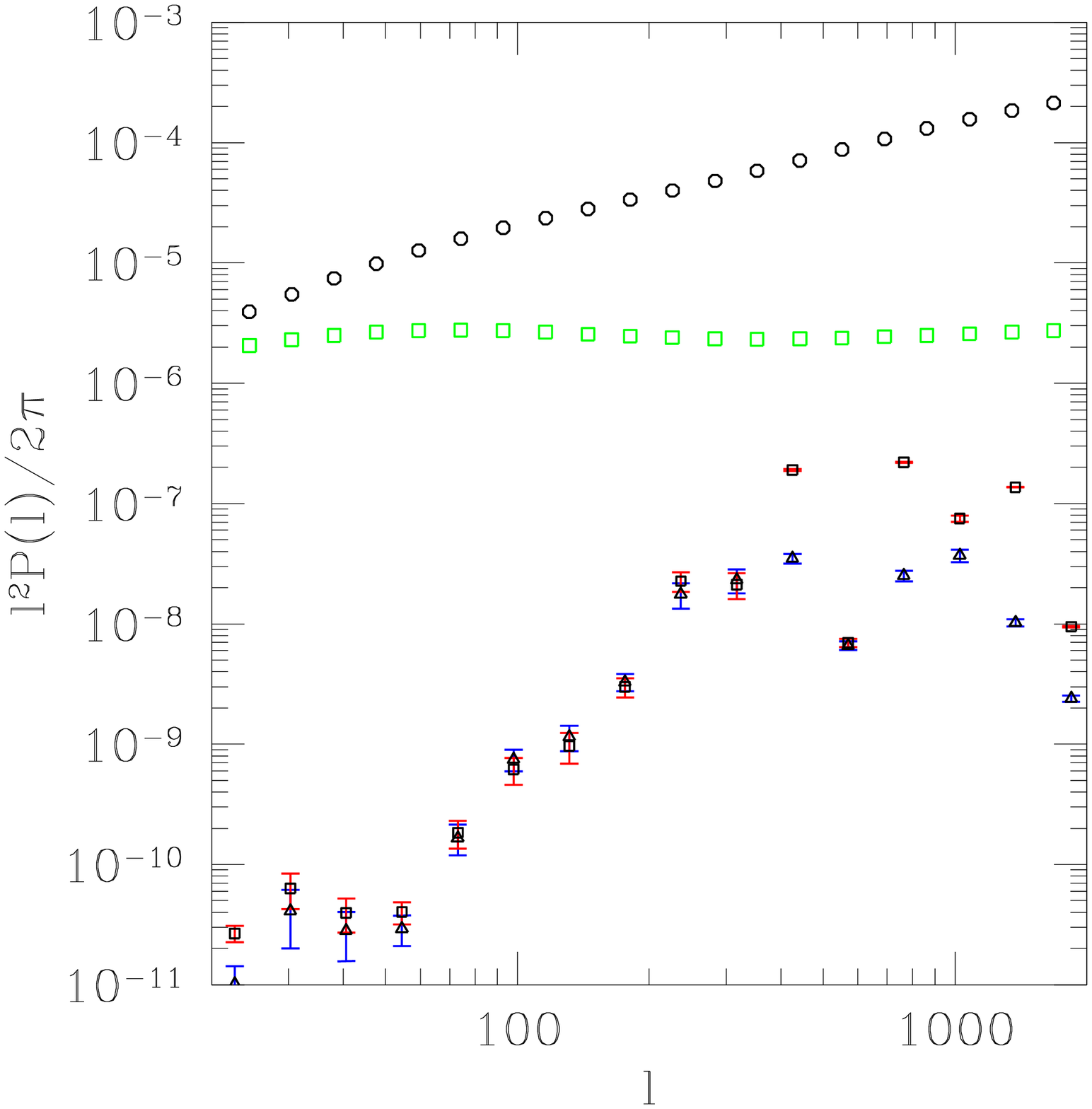}
  \caption{\small Power spectra of the lensing signal and expected PSF
    anisotropy contamination for the SNAP telescope.  Thermal effects
    and vibration in the structure of the telescope lead to time
    variation in the PSF anisotropy.  This residual power would lead
    to an additive systematic error in the lensing measurement.
    Subtracting the static pattern (see text) potentially provides a
    reduction in PSF power of as much as an order of magnitude.  The
    red curve (squares) is the time-varying power due to thermal
    effects, and has not had the static PSF pattern subtracted off,
    whereas the blue curve (triangles) is the residual power after the
    time invariant pattern is subtracted (as it is expected to be
    measured accurately). The statistical errors for a 1000 square
    degree survey are shown by the green squares, along with the power
    spectrum amplitude of the shear signal from simulations (circles).
    The statistical error is projected to be larger by $> 1$ orders
    of magnitude than the time-varying residual PSF signal.  The much
    smaller residual power from structural vibrations is about two
    orders of magnitude below the bottom of the plot.
}
  \label{compare_resid_rayt}
\end{figure}

The systematic error contribution to the lensing shear power spectrum
can be obtained by Fourier transforming of the gridded PSF ellipticity
maps described above.  The power spectrum due to the thermal drift is
shown in Figure \ref{compare_resid_rayt}, along with the power
spectrum due to vibrations, which has a much smaller amplitude. The
power spectrum due to jitter is smaller still, and is below the scale
plotted.  For the thermal drift, the full contribution to the power
spectrum, and the smaller contribution due the residual shear after
the static pattern is subtracted are shown.  This subtraction is
important; for real data it is expected that the static PSF can be
measured accurately by interpolating from stars in the image \citep{Ja04}.

\begin{figure}[t]
  \epsscale{1.0}
  \plotone{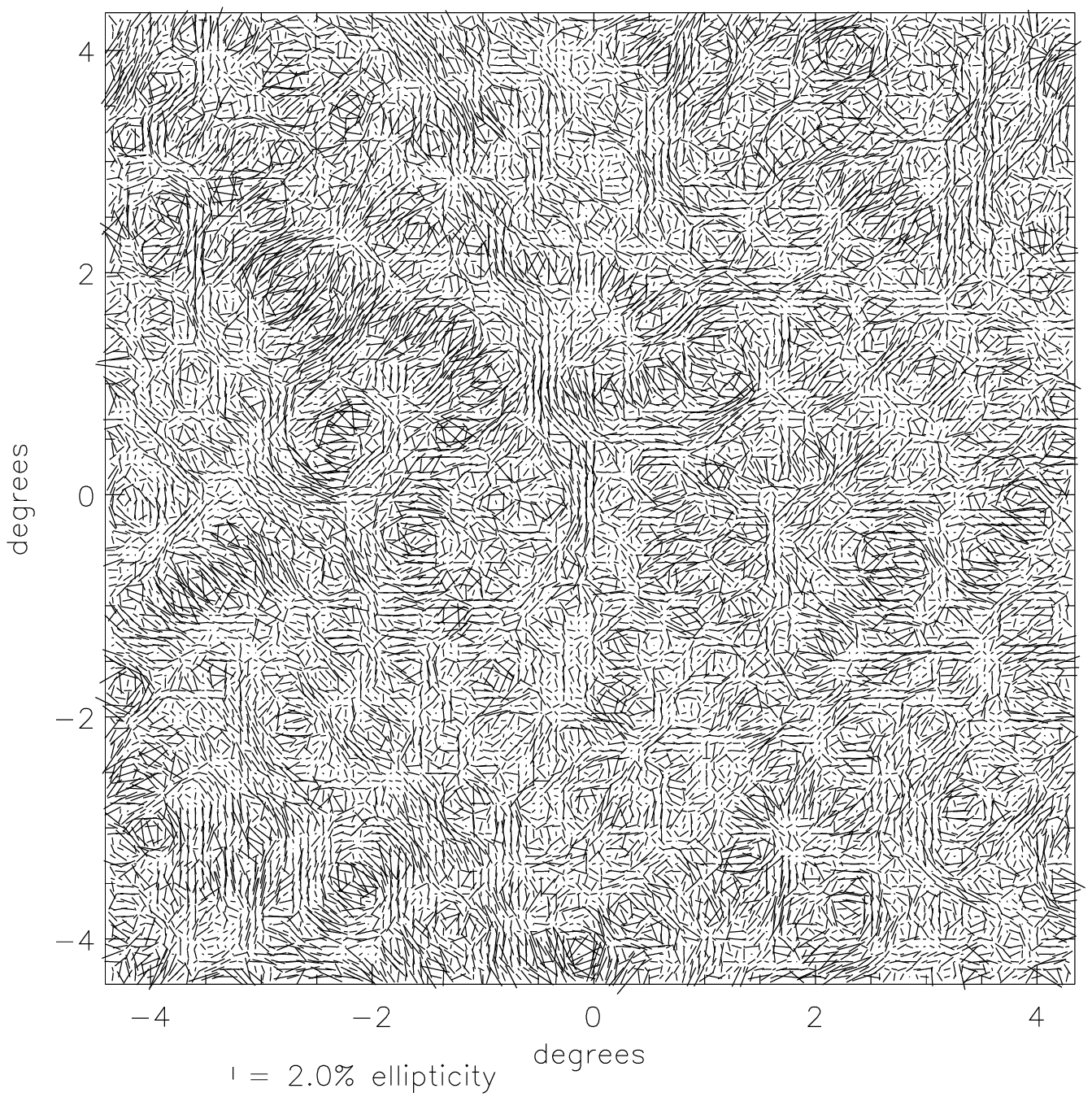}
  \caption{\small Ray-tracing through N-body simulation data provides
    a way to estimate the importance of the time-varying SNAP PSF.
    The power spectrum of the ensemble of simulations is the green
    curve in Figure~\ref{compare_resid_rayt}.}
  \label{rayt_whiskers}
\end{figure}

In order to compare the effects of PSF anisotropy with the lensing
signal, we generated some simulated shear fields via ray-tracing
through N-body simulation boxes \citep{Heitmann_Nbody}.  The
simulations were performed in a $\rm \Lambda CDM$ cosmology with a
comoving box size of 500~$\rm h^{-1} Mpc$, and the sources were set at
a redshift of $z=2$.  Figure~\ref{rayt_whiskers} shows a whisker plot
of one of the realizations; the RMS shear in the field shown is about
$2\%$.  Figure~\ref{compare_resid_rayt} shows that the power spectrum
of the time-varying residual of the PSF is typically smaller than the
lensing signal by three to four orders of magnitude. It is also much
smaller than the statistical errors expected for a thousand square
degree survey, also shown in the plot. These include sample variance
and the shot noise contribution of galaxies with an RMS ellipticity of
$\sigma_\epsilon = 0.4$ (both components combined) and number density
$n_g=100$ per square arcminute.  Thus for the standard parameters of
the telescope, we have shown that the power spectrum of the residual
error due to PSF anisotropy makes a negligible contribution to the
lensing power spectrum.

\section{Discussion}

We have estimated the effect of three time varying components of PSF anisotropy
to the weak lensing power spectrum for the optical design of the SNAP telescope. 
The time variation of the three effects considered, thermal drift, guider jitter
and structural vibrations, leads to spatial patterns in the shear maps 
inferred from the measured ellipticities of galaxies. We find that for the 
current best estimates of the amplitude of these three effects, the contribution 
to the lensing power spectrum is much smaller than the expected statistical 
errors (after subtracting the static component of the PSF pattern). 
The systematic error contribution due to PSF anisotropy is therefore 
negligible, excluding catastrophic failures, revised estimates of the effects 
we have modeled, or other effects not included in this study. 

It would further be of interest to compute the bispectrum of the PSF
pattern and compare it to the signal for different triangle
configurations.  The bispectrum carries important cosmological
information, so it is important to test how it will be affected by
systematic errors.  Preliminary results for special configurations
suggest that the contribution to the bispectrum is also much smaller
than the expected signal. We will further include worst-case sources
of misalignment and catastrophic events such as an extended power
outage. Finally, we note that even in the presence of large systematic
errors, the estimate of cosmological parameters may be compromised
only to a limited extent, as the lensing signal for multiple auto and
cross-spectra will in general scale differently with these parameters
\citep{Huterer}.

\acknowledgements

We thank Katrin Heitmann, Salman Habib, and Derek Dolney for help with
N-body simulation data.  We acknowledge useful suggestions from Mike
Sholl, Michael Levi, and Saul Perlmutter.

\end{document}